# Weakly Coupled Motion of Individual Layers in Ferromagnetic Resonance

D.A. Arena, E. Vescovo, C.-C. Kao
*National Synchrotron Light Source, Brookhaven National Lab, Upton, NY*

Y. Guan, W.E. Bailey
*Materials Science Program, Department of Applied Physics,
Columbia University, New York, NY*

## Abstract

We demonstrate a layer- and time-resolved measurement of ferromagnetic resonance (FMR) in a $Ni_{81}Fe_{19}$ / Cu / $Co_{93}Zr_7$ trilayer structure. Time-resolved x-ray magnetic circular dichroism has been developed in transmission, with resonant field excitation at a FMR frequency of 2.3 GHz. Small-angle (to 0.2°), time-domain magnetization precession could be observed directly, and resolved to individual layers through elemental contrast at Ni, Fe, and Co edges. The phase sensitivity allowed direct measurement of relative phase lags in the precession oscillations of individual elements and layers. A weak ferromagnetic coupling, difficult to ascertain in conventional FMR measurements, is revealed in the phase and amplitude response of individual layers across resonance.

## INTRODUCTION

The dynamic properties of soft ferromagnets (FM) play an obviously important role in determining the characteristics of many modern magneto-electronic devices, from giant magneto-resistive (GMR) read heads in hard disk drives to magnetic tunnel junctions and other advanced "spintronic" devices [1,2]. Precessional dynamics at 1-10 GHz determine the high-speed response of ferromagnetic heterostructures, and present a fundamental limit to increasing data rates in magnetic information storage [3,4]. FM materials in such structures are often complex, multilayer systems comprised of metallic alloys and other compounds. The interplay between the various layers, or between elemental moments in a single layer, can have a dramatic effect on the high-speed response. It is also the motivation for developing new measurement techniques that can separate the dynamic behavior of these complex structures in a layer-by-layer or element-by-element basis [5-13]

Ferromagnetic resonance (FMR) remains a technique of choice for measuring fundamental quantities in precessional dynamics. In FMR, a microwave-frequency RF magnetic field excites motion of the ferromagnetic moments, most strongly where the microwave driving frequency matches the resonant frequency of magnetization precession. The conical precessional motion of **M** is indicated schematically in Fig 1(a); for thin films magnetized in plane, dipolar fields compress the orbit into the film plane and the cone becomes strongly elliptical. The response is measured by microwave absorption; the position and width of the resonant absorption provide quantitative information on the resonance and relaxation frequencies, respectively.

-

The motion of the precession can be parameterized with two variables: the cone angle $\theta$ and the phase of the oscillation $\phi$. In the case of two weakly coupled systems, a phase lag, $\Delta\phi$, can develop between the two precessing magnetization vectors (Figs. 1(b) & 1(c)). Sinusoidal motion of the transverse component of **M** is expected during precession. As the system is swept through resonance, by either varying the microwave frequency or modifying the resonant frequency via an external bias field, this motion should go through a maximum in amplitude and a 180° shift in phase. For different, uncoupled FM layers with separate resonances, each responds independently while swept through resonance. For near-identical layers coupled ferromagnetically, resonance frequencies are split into lower-frequency acoustic modes (in-phase for $M_1$ and $M_2$) and higher-frequency optical modes (out-of-phase for $M_1$ and $M_2$). Two weakly coupled layers with separated resonances will show mixed behavior. [14]

It is worth noting that *layer-resolved* measurements of FMR are particularly relevant to magnetic multilayer structures where the coupling can be varied over a considerable range by judicious selection of non-magnetic (NM) spacer materials, thickness, and interface roughness. For example, indirect exchange coupling (IEC) of FM layers through a NM spacer, as in $FM_1/NM/FM_2$, has been investigated since the beginnings of interest in magnetoelectronics [15]. Conventional FMR has provided quantitative estimates of parameters such as the magnitude and sign of the ferromagnetic/antiferromagnetic coupling constant $A_{ex}$ between the FM layers through splitting of the coupled resonance frequencies[14]. However, in coupled layers, no measurement technique could directly separate the response of $FM_1$ from $FM_2$; visible magnetooptical and electronic measurements combine the response of both layers, and separation of the motion is achieved only through models.

In this article, we demonstrate a element- and time-resolved measurement of ferromagnetic resonance (ETR-FMR) that distinguishes precession in individual layers. The phase and amplitude of the response of individual layers is revealed using time-resolved x-ray magnetic circular dichroism (TR-XMCD), developed in resonant-field pumping and with a transmission geometry. The novel configuration enables orders of magnitude improvement in the resolution of precessional amplitude (to 0.2°) and phase (to 5 ps, or 5° at 2.3 GHz), respectively. Weak ferromagnetic coupling between a $Ni_{81}Fe_{19}$ and $Co_{93}Zr_7$ layers in a $Ni_{81}Fe_{19}/Cu/Co_{93}Zr_7$ "pseudo-spin valve" trilayer is observed clearly through the amplitude and phase responses of the individual layers, measured separately at the (Ni,Fe) and Co x-ray absorption edges. The weak coupling level (0.01 erg/cm$^2$, or ~ 6 Oe) is not easily detected through complimentary microwave absorption measurement.

**THEORY**

We begin our modelling of a weakly coupled FM trilayer system starting from the Landau-Lifschitz (LL) equation of motion for magnetization dynamics [16], given in SI as:

$$\frac{d\mathbf{M}^{(i)}}{dt} = -\mu_0 |\gamma^{(i)}| \left(\mathbf{M}^{(i)} \times \mathbf{H}_{\text{eff}}^{(i)}\right) - \frac{\lambda^{(i)}}{|M_s^{(i)}|^2} \left[\mathbf{M}^{(i)} \times \left(\mathbf{M}^{(i)} \times \mathbf{H}_{\text{eff}}^{(i)}\right)\right] \qquad (1)$$

where the superscript i indexes FM layer 1 or 2 in the trilayer structure. $\mathbf{M}^{(i)}(t)$ is the magetization with saturation value $M_s^{(i)}$, $\gamma^{(i)}$ is the gyromagnetic ratio, $\lambda^{(i)}$ is the LL relaxation rate in $sec^{-1}$ and $\mathbf{H}_{\text{eff}}^{(i)}$ is the effective field in the $i^{th}$ FM layer.

-

In our trilayer system, $\mathbf{H}_{\text{eff}}^{(i)}$ consists of two external fields, $\mathbf{H_B}$ and $\mathbf{h}_{\text{rf}}(t)$, and three internal fields related to the magnetic properties of the films:

$$\mathbf{H}_{\text{eff}}^{(i)} = \mathbf{H_B} + \mathbf{h}_{\text{rf}}(t) + \mathbf{H}_{\text{D}}^{(i)} + \mathbf{H}_{\text{k}}^{(i)} + \mathbf{H}_{\text{ex}}^{(i)} \qquad (2)$$

where (1) external DC bias field $\mathbf{H_B}$, (2) external rf microwave field $\mathbf{h}_{\text{rf}}(t)$, (3) demagnetization field $\mathbf{H}_{\text{D}}^{(i)}$, (4) effective magneto-crystalline anisotropy field $\mathbf{H}_{\text{k}}^{(i)}$, and (5) an inter-layer coupling field $\mathbf{H}_{\text{ex}}^{(i)}$.

As shown in Fig. 1(d), we take the ferromagnetic trilayer structure to lie in the x-y plane, with z normally directed. The external DC bias field $\mathbf{H_B}$ is taken to be along the x-axis, the rf microwave driving field $\mathbf{h}_{\text{rf}}(t)$ is along the y-axis, the effective magneto-crystalline anisotropy field $\mathbf{H}_{\text{k}}^{(i)}$ is along the x-axis, and the demagnetization field $\mathbf{H}_{\text{D}}^{(i)}$ is along the z-axis.

To arrive an expression for $\mathbf{H}_{\text{ex}}^{(i)}$, we consider the energy per unit area of the coupling between the FM layers, which can be expressed as:

$$E_{ex} = -A_{ex} \frac{\mathbf{M}^{(i)} \cdot \mathbf{M}^{(j)}}{M_s^{(i)} M_s^{(j)}} \qquad (3)$$

where the coupling constant $A_{ex}$ has units of energy/unit area and is assumed to be positive for parallel, ferromagnetic coupling between the FM layers and negative for antiparallel, antiferromagnetic coupling between the FM layers. The exchange-coupling field can then be expressed as a derivative of Eq. (3): $\mathbf{H}_{\text{ex}}^{(i)} = -(1/\mu_0 t^{(i)})(\partial E_{ex}/\partial \mathbf{M}^{(i)})$, with the result as:

$$\mathbf{H}_{\text{ex}}^{(i)} = \frac{A_{ex}}{\mu_0 M_s^{(i)} t^{(i)}} \cdot \frac{\mathbf{M}^{(j)}}{M_s^{(j)}} = \frac{A_{ex}}{\mu_0 M_s^{(i)} t^{(i)}} \cdot \mathbf{m}^{(j)} \qquad (4)$$

where $t^{(i)}$ is the thickness of the ith FM layer and the superscript j indexes the other FM layer in the trilayer structure. We will consider only a small deviation of the magnetization $\mathbf{M}^{(i)}$ (i=1,2) from its equilibrium position $M_s^{(i)} \hat{\mathbf{e}}_{r,i}$:

$$\mathbf{M}^{(i)} \approx M_s^{(i)} \hat{\mathbf{e}}_{r,i} + m_{\theta i} \hat{\mathbf{e}}_{\theta i} + m_{\phi i} \hat{\mathbf{e}}_{\phi i} \tag{5}$$

where $m_{\theta i} = M_s^{(i)} \Theta_i$ and $m_{\phi i} = M_s^{(i)} \sin(\theta_i) \Phi_i$ are small deviations of the magnetizations along the $\theta_i$ and $\phi_i$ directions, respectively. $\theta_i$ is the magnetization angle away from the z-axis; $\phi_i$ is the magnetization angle away from the x-axis in the x,y plane. $\Theta_i$ is expressed in terms of $\Theta_i = \theta_i - 90°$ so that $\Theta_i = 0$ has magnetization in-plane. Assuming in-plane magnetization at equilibrium for both FM layers, $\mathbf{m}^{(i)}$ (i = 1,2) in Eq. (4) can be expressed as:

$$\mathbf{m}^{(i)} \approx \hat{\mathbf{e}}_\mathbf{x} + \Phi_i \hat{\mathbf{e}}_\mathbf{y} - \Theta_i \hat{\mathbf{e}}_\mathbf{z} \tag{6}$$

and the demagnetization field $\mathbf{H}_\mathbf{D}^{(i)}$ (i = 1,2) can be written as:

$$\mathbf{H}_\mathbf{D}^{(i)} \approx M_s^{(i)} \Theta_i \hat{\mathbf{e}}_\mathbf{z} \tag{7}$$

Under above assumptions, the linearized Landau-Lifshitz equations of motion for small precessional angles ($\Theta_i$ and $\Phi_i$ for i=1, 2) can be obtained by expanding Eq. (1) in spherical coordinates and retaining only terms to the first order of $\Theta_i$ and $\Phi_i$, shown as the following:

$$\frac{d\Theta_i}{dt} \approx \frac{\lambda^{(i)}}{M_s^{(i)}} H_\theta^{(i)} + \mu_0 |\gamma^{(i)}| H_\phi^{(i)} \tag{8}$$

$$\frac{d\Phi_i}{dt} \approx -\mu_0 |\gamma^{(i)}| \left( H_\theta^{(i)} + \frac{\lambda^{(i)}}{M_s^{(i)}} H_\phi^{(i)} \right) \tag{9}$$

where $H_\theta^{(i)}$ and $H_\phi^{(i)}$ can be expressed in terms of the Cartesian fields $H_x^{(i)}$, $H_y^{(i)}$, and $H_z^{(i)}$:

$$H_\theta^{(i)} \approx -\Theta_i H_x^{(i)} - H_z^{(i)} \tag{10}$$

$$H_\phi^{(i)} \approx -\Phi_i H_x^{(i)} + H_y^{(i)} \tag{11}$$

Being more explicit here, we can write the Cartesian fields $H_x^{(i)}$, $H_y^{(i)}$, and $H_z^{(i)}$ as:

$$H_x^{(i)} = H_B + H_k^{(i)} + \frac{A_{ex}}{\mu_0 M_s^{(i)} t^{(i)}} \tag{12}$$

$$H_y^{(i)} = h_{rf}(t) + \frac{A_{ex}}{\mu_0 M_s^{(i)} t^{(i)}} \Phi_i \tag{13}$$

$$H_z^{(i)} = M_s^{(i)} \Theta_i - \frac{A_{ex}}{\mu_0 M_s^{(i)} t^{(i)}} \Theta_j \tag{14}$$

where j = 1(2) if I = 2(1).

Finally, if a harmonic form for the external driving field $h_{rf}(t)$ (i.e. $h_{rf}(t) = h_{rf,0} \exp(i\omega t)$) is assumed, as is the case in force resonance, the linearized Landau-Lifshitz equations of motion for small precessional ($\Theta_i$ and $\Phi_i$ for i=1, 2) can be written in matrix form as:

$$\frac{d}{dt} \begin{bmatrix} \Theta_1 \\ \Phi_1 \\ \Theta_2 \\ \Phi_2 \end{bmatrix} \approx \mathbf{A} \begin{bmatrix} \Theta_1 \\ \Phi_1 \\ \Theta_2 \\ \Phi_2 \end{bmatrix} + \mathbf{g(t)} \tag{15}$$

-

where:

$$\mathbf{A} = \begin{bmatrix} -\lambda^{(1)} & -\mu_0|\gamma^{(1)}|\widetilde{H}^{(1)} & \frac{\lambda^{(1)}}{M_s^{(1)}}H_{ex}^{(1)} & \mu_0|\gamma^{(1)}|H_{ex}^{(1)} \\ \mu_0|\gamma^{(1)}|(M_s^{(1)}+\widetilde{H}^{(1)}) & -\frac{\lambda^{(1)}}{M_s^{(1)}}\widetilde{H}^{(1)} & -\mu_0|\gamma^{(1)}|H_{ex}^{(1)} & \frac{\lambda^{(1)}}{M_s^{(1)}}H_{ex}^{(1)} \\ \frac{\lambda^{(2)}}{M_s^{(2)}}H_{ex}^{(2)} & \mu_0|\gamma^{(2)}|H_{ex}^{(2)} & -\lambda^{(2)} & -\mu_0|\gamma^{(2)}|\widetilde{H}^{(2)} \\ -\mu_0|\gamma^{(2)}|H_{ex}^{(2)} & \frac{\lambda^{(2)}}{M_s^{(2)}}H_{ex}^{(2)} & \mu_0|\gamma^{(2)}|(M_s^{(2)}+\widetilde{H}^{(2)}) & -\frac{\lambda^{(2)}}{M_s^{(2)}}\widetilde{H}^{(2)} \end{bmatrix}, \quad (16)$$

$$\mathbf{g(t)} = \begin{bmatrix} \mu_0|\gamma^{(1)}| \\ \frac{\lambda^{(1)}}{M_s^{(1)}} \\ \mu_0|\gamma^{(2)}| \\ \frac{\lambda^{(2)}}{M_s^{(2)}} \end{bmatrix} h_{rf}(t) \quad (17)$$

and

$$H_{ex}^{(i)} = \frac{A_{ex}}{\mu_0 M_s^{(i)} t^{(i)}} \quad (18)$$

$$\widetilde{H}^{(i)} = H_B + H_k^{(i)} + H_{ex}^{(i)} \ll M_s^{(i)} \quad (19)$$

$$h_{rf}(t) = h_{rf,0}\exp(i\omega t) \quad (20)$$

**EXPERIMENT**

The XMCD and ETR-FMR measurements were carried at the circularly polarized soft x-ray beamline 4-ID-C of the Advanced Photon Source (APS) at Argonne National Laboratory. Fig. 2(a) presents a block diagram of apparatus for in-situ measurements of conventional FMR and ETR-FMR. The signal from the APS photon bunch clock is first

-

directed through a variable delay and then to a phase-locked frequency synthesizer that upconverts the 88 MHz signal to 2.3 GHz. The amplified output is then directed into a hollow broadband rf resonator to excite the uniform precession of the magnetic thin film sample inside. Two orthogonal sets of Helmholtz coils provide the vertical bias field $H_B$ or the horizontal transverse field $H_T$. Conventional in-situ FMR measurements are made by varying $H_B$ and detecting the reflected power in the rf circuit via use of directional coupler and a microwave diode. Fig. 2(b) presents a top view of experimental geometry. The sample is rotated 38° with respect to the incident photon direction, which is parallel to the transverse magnetic field $H_T$. The transmitted x-ray photons are detected using a standard soft x-ray photodiode.

The trilayer sample, $Ni_{81}Fe_{19}$ (25 nm)/ Cu (20 nm)/ $Co_{93}Zr_7$ (25 nm)/ Cu cap (5 nm), was grown via UHV magnetron sputtering at a base pressure of $4\times10^{-9}$ Torr. The substrate used was a commercially available $Si_3N_4$ membrane, 100 nm in thickness. At the transition metal edges of interest, transmission through the substrate is greater than 80%. For ETR-FMR and standard FMR measurements, the sample was placed inside the hollow broadband rf resonator.

Static XMCD spectra of Fe, Ni and Co were measured in transmission mode, with strong dichroism signals easily visible on the $L_3$ and $L_2$ edges of all FM elements (Fig. 3). Element specific XMCD hysteresis loops, also shown in Fig. 3, were taken as a function of the transverse field $H_T$ to obtain a calibration for rotational magnetization angle $\phi$, by tuning the photon energy to the peak XMCD signal (arrows, $h\nu$ = 707.5 eV for Fe, 851.5 eV for Ni, and 778 eV for Co). The Fe and Ni loops indicates that the $Ni_{81}Fe_{19}$ layer switches at a coercive field of 5 Oe; the $Co_{93}Zr_7$ layer exhibits a larger coercive field of 7

Oe. All the three XMCD spectra show a negative contribution at the $L_3$ edge and thus any static coupling between the FM layers is taken to favor parallel alignment of the layers. Also, the thin film sample presents an in-plane anisotropy and the saturation values of each hysteresis loop are taken to be ±90°.

ETR-FMR timing scans were conducted at a fixed continuous-wave (CW) rf frequency of 2.3 GHz; the CW rf excitation produced a time-varying magnetic field ($\mathbf{h_{rf}}$) that was in the film plane and transverse to the time-averaged magnetization direction (Fig. 1(d)). For ETR-FMR, $\mathbf{h_{rf}}$ was phase-locked with the APS photon bunch clock (88 MHz) and thus the instantaneous position of $\mathbf{M_{Fe,Ni}}$ or $\mathbf{M_{Co}}$ was sampled stroboscopically with the photon pulses once every 26 precession cycles. ETR-FMR timing scans were recorded by tuning the photon energy to the transition metal edge of interest (arrows, Fig. 3), setting the bias field $\mathbf{H_B}$ to the desired point on the conventional in-situ FMR resonance curve (Fig. 4(a)), and sweeping out the variable delay between the 2.3 GHz CW rf excitation and the APS bunch clock.

**RESULTS AND DISCUSSION**

An i*n-situ*, conventional FMR spectrum using the 2.3 GHz, phase-locked CW excitation is presented in Fig 4(a). A strong resonance, identified by the zero crossing point, is readily apparent at $H_B = 40$ Oe. A much weaker inflection is observed near zero bias. In addition, Fig. 4(b) presents measurements of *ex-situ*, variable frequency conventional FMR. Plotted in Fig. 4(b) are the resonance field ($H_{B,res}$) at the square of the variable excitation frequencies ($f_p^2$). At higher frequencies than the 2.3 GHz used for ETR-FMR measurements (indicated by the horizontal dashed dotted line), the weak

-

resonance near $H_B = 0$ Oe in Fig. 4(a) increases in intensity and is more clearly defined. Two branches are observed in the data in Fig. 4(b), which correspond primarily to separate resonances in the $Ni_{81}Fe_{19}$ and $Co_{93}Zr_7$ layers. The LL equations for magnetization dynamics can model satisfactorily the data in the absence of coupling between the layers (solid green lines, Fig. 4(b)) or with a very weak FM coupling (dashed magenta lines, Fig. 4(b)).

ETR-FMR timing scans for the $Ni_{81}Fe_{19}$ / Cu / $Co_{93}Zr_7$ trilayer were acquired at the field values indicated by the black arrows in Fig. 4(a). Results are exhibited in Fig. 5, where the timing scans have been offset for clarity All of the data in Fig. 5 exhibit a strong sinusoidal dependence with time delay and therefore the timing scans in the figure can be fit by simple sinusoidal functions (solid black lines). The amplitudes and phases determined by these fits have a high degree of confidence. Precession amplitudes as small as 0.2° have been measured and the estimated errors are on the order of 0.02° (one standard deviation). The ETR-FMR data also allow for precise determination of the phase of the oscillations; the resolution of the phase determination was 2°~ 5°. At 2.3 GHz, this translates to an overall timing resolution of 2 ~ 6 ps. The timing resolution achieved in ETR-FMR measurements represents a tenfold to hundredfold improvement over previous tr-XMCD results using pulsed excitations [5]. Significantly, the phase sensitivity of ETR-FMR enables a time resolution an order magnitude smaller than the intrinsic bunch width of the x-ray photons (~60 ps).

In Fig. 5, the top panel presents the time delay scans at different values of $H_B$ for the Co moment oscillation in the $Co_{93}Zr_7$ layer. The bottom panel presents similar data for Fe in the $Ni_{81}Fe_{19}$ film. A representative timing scan for Ni is also presented for $H_B = 40$

-

Oe. The moments of the Fe and Ni in the $Ni_{81}Fe_{19}$ layer are seen to precess in unison within the instrumental resolution of ~0.2° in amplitude and ~2 ps in phase. The observation of coupled motion between the Fe and Ni moments is a signature of strong exchange coupling within the $Ni_{81}Fe_{19}$ layer that binds the moments of the two FM elements in lockstep, which is consistent with previous TR-XMCD results [5].

The phase and amplitude parameters extracted from the time delay scans at different bias fields (for $H_B$ = 14, 21, 27, 33, 40 and 46 Oe) are presented for Co and Fe in Fig. 6(a) (amplitude) and Fig. 6(b) (phase). In contrast to the identical precession of the Fe and Ni moments within the $Ni_{81}Fe_{19}$ layer, the dynamics *between* FM layers is very different. The clearest trend in Fig. 6(a) is the dramatic increase in amplitude of the Fe oscillations as $H_B$ is increased from 14 Oe (0.34°) to 40 Oe (1.35°), followed by a decline at 46 Oe (1.13°). In contrast, the Co amplitude decreases monotonically. A comparison with Fig. 4(a) indicates that the rapid growth and subsequent decline in the amplitude of the Fe oscillations is a consequence of driving the system through a resonance. Based on the 2.3 GHz FMR data (Fig. 4(a)) and the FMR measurements at higher frequency (Fig. 4(b)), the increase in the Co oscillations at low values of $H_B$ is consistent with a weak resonance in the $Co_{93}Zr_7$ layer near zero bias.

The changes in the phase of the oscillations in the two layers (Fig. 6b) as the system is forced into resonance exhibit a corresponding behavior. From $H_B$ = 14 Oe through $H_B$ = 46 Oe the Fe (and Ni) phase undergoes a dramatic change of ~100°. This is consistent with the expected phase shift of ~180° upon going through a resonance. In contrast, the phase of the Co signal remains largely unaffected. Also, note that the precession of $\mathbf{M_{Fe,Ni}}$ and $\mathbf{M_{Co}}$ differs in phase by ~120° for the lowest value of $H_B$ measured (14 Oe),

-

and that the phase difference decreases to less than 10° at the highest value of $H_B$ measured (46 Oe).

The main features of the changes in the amplitude and phase of precession in $\mathbf{M_{Fe,Ni}}$ and $\mathbf{M_{Co}}$ are qualitatively consistent with independent resonances in the $Ni_{81}Fe_{19}$ and $Co_{93}Zr_7$ layers. Closer inspection, however, reveals a weak coupling between the two layers. Also shown in Fig. 6 are the calculated values for amplitude and phase of the oscillations based on numerical solutions from the two-particle linearized LL model (coupled linearized LL equations of motion for in $\mathbf{M_{Fe,Ni}}$ and $\mathbf{M_{Co}}$). The coupling is parameterized as an effective exchange energy per unit area, $A_{ex}$ in Eq. 4. Results assuming no coupling between the layers (*i.e.* $A_{ex} = 0$, dashed lines) and weak coupling ($A_{ex} = 0.01$ ergs / cm$^2$, solid lines) are shown in the figures. For the amplitudes of the precession, the assumption of no coupling between the layers overestimates the resonance field $H_{B,res}$ for the main resonance in the $Ni_{81}Fe_{19}$ layer which is observed near 40 Oe and also predicts a narrower width for the amplitude than is observed; an assumption of weak coupling more closely matches the observed motion of $\mathbf{M_{Fe,Ni}}$. For the $Co_{93}Zr_7$ layer, weak coupling again provides better agreement with the data as an assumption of uncoupled layers overestimates the amplitude of the precession cone for Co. In the case of the phase of the oscillations, an assumption of independent layers ($A_{ex} = 0$, dashed lines) predicts a more gradual decline in the phase of the $Ni_{81}Fe_{19}$ layer than is observed, and also does not account for the upturn in the phase of the $Co_{93}Zr_7$ layer at high values of $H_B$. Again, an assumption of weak coupling ($A_{ex} = 0.01$ ergs / cm$^2$, solid lines) provides better agreement with the measurements. By observing the motion, in both

-

amplitude and phase, of individual layers, ETR-FMR can thus reveal weak coupling much more directly than the conventional, variable frequency FMR presented in Fig. 4(b).

The most likely causes of the weak coupling between $\mathbf{M_{Fe,Ni}}$ and $\mathbf{M_{Co}}$ are the previously mentioned IEC mechanism or a *Neél* ("orange-peel") type dipolar coupling [17,18]. *Neél* coupling is more probable as IEC has a considerably shorter interaction range [19] and our 20 nm Cu spacer layer is rather thick. Both mechanisms are static interactions that nonetheless can affect the dynamic response of a coupled system. Recently, *dynamic* coupling mechanisms have also been proposed. For example, Tserkovnyak *et al.* propose a relatively long range, spin pumping mechanism in FM / NM / FM trilayers where precession in one FM layer affects the damping in the other FM layer [20,21]. Conventional FMR has been used to examine these systems and effects arising from spin-pumping have been detected in resonance line widths [22,23]. ETR-FMR, however, would be able to measure such effects directly by looking at FM layers individually.

ETR-FMR is inherently a core-level, local-probe technique and as such is well-positioned to illuminate a number of unresolved issues in magnetization dynamics. For example, lagged-response models developed to explain damping mechanisms in magnetization dynamics typically incorporate energy transfer between different reservoirs in a magnetic system; this energy transfer is made manifest via phase lags between different constituents [24]. ETR-FMR, which can observe the magnetic dynamics of individual elements, can examine directly such phase lags in the response of a subsystem. Also, as ETR-FMR is based on the well-developed XMCD technique, *spectroscopic* investigations of dynamics, such as dynamic variations in the orbit-to-spin ratio of magnetic moments or induced magnetism in non-magnetic layers during

-

precession, are now feasible. Such measurements should provide valuable inputs to emerging first-principles based calculations that seek to move beyond the phenomenological approaches inherent in the LL theory and its extensions.

**CONCLUSION**

A layer- and time-resolved ferromagnetic resonance, in a "pseudo-spin valve" structure of $Ni_{81}Fe_{19}$ / Cu / $Co_{93}Zr_7$ has been measured by time-resolved XMCD in transmission mode synchronized with CW microwave excitation at 2.3 GHz. The phase and amplitude of driven FMR precession have been observed magneto-optically, with a very high temporal and rotational sensitivity. More importantly, a weak ferromagnetic coupling between the two FM layers has been revealed much more transparently than any conventional FMR measurements.

**ACKNOWLEDGEMENTS**

The authors thank David J. Keavney (APS) for beamline support. This work was partially supported by the Army Research Office with Grant No. ARO-43986-MS-YIP, and the National Science Foundation with Grant No. NSF-DMR-02-39724. Use of the Advanced Photon Source was supported by the U.S. Department of Energy, Office of Science, Office of Basic Energy Sciences, under Contract No. W-31-109-Eng-38.

-

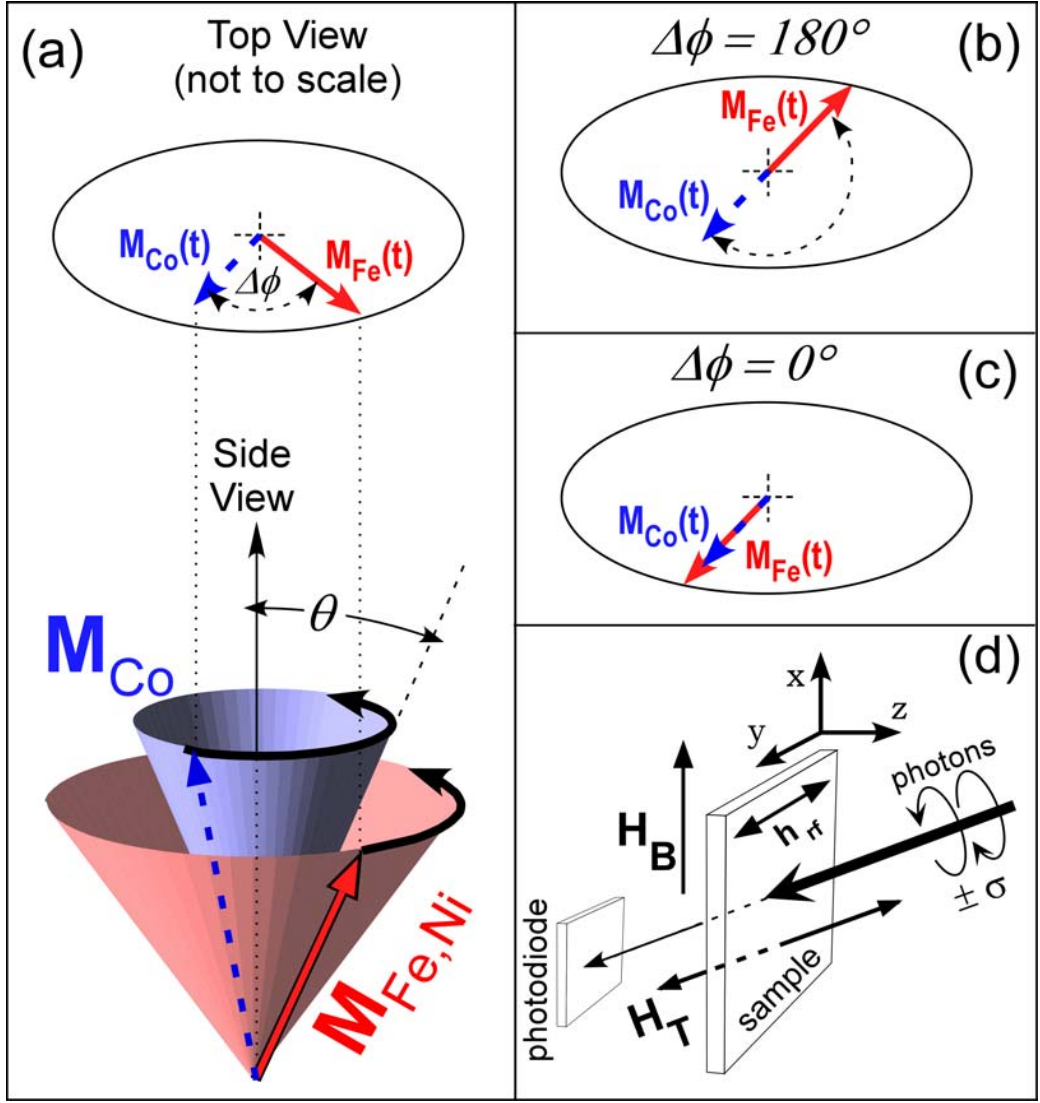

FIG. 1 (color online): (a) side view of precession cones for $\mathbf{M_{Fe,Ni}}$ and $\mathbf{M_{Co}}$ showing the cone angle $\theta$ and top view of phase difference, $\Delta\phi$, between the two vectors. (b) top view indicating out-of-phase difference. (c) top view indicating in-phase condition. (d) Diagram of measurement geometry. The sample is oriented vertically in the x-y plane with the sample normal parallel to z. Photons are incident along the horizontal (in the y-z plane), rotated 38° away from the z-axis. Orthogonal Helmholtz coils provide a magnetic field parallel to the film plane along x (bias field $\mathbf{H_B}$) or parallel to the incident photon direction (transverse field $\mathbf{H_T}$). Transmitted photons are detected with a photodiode.

-

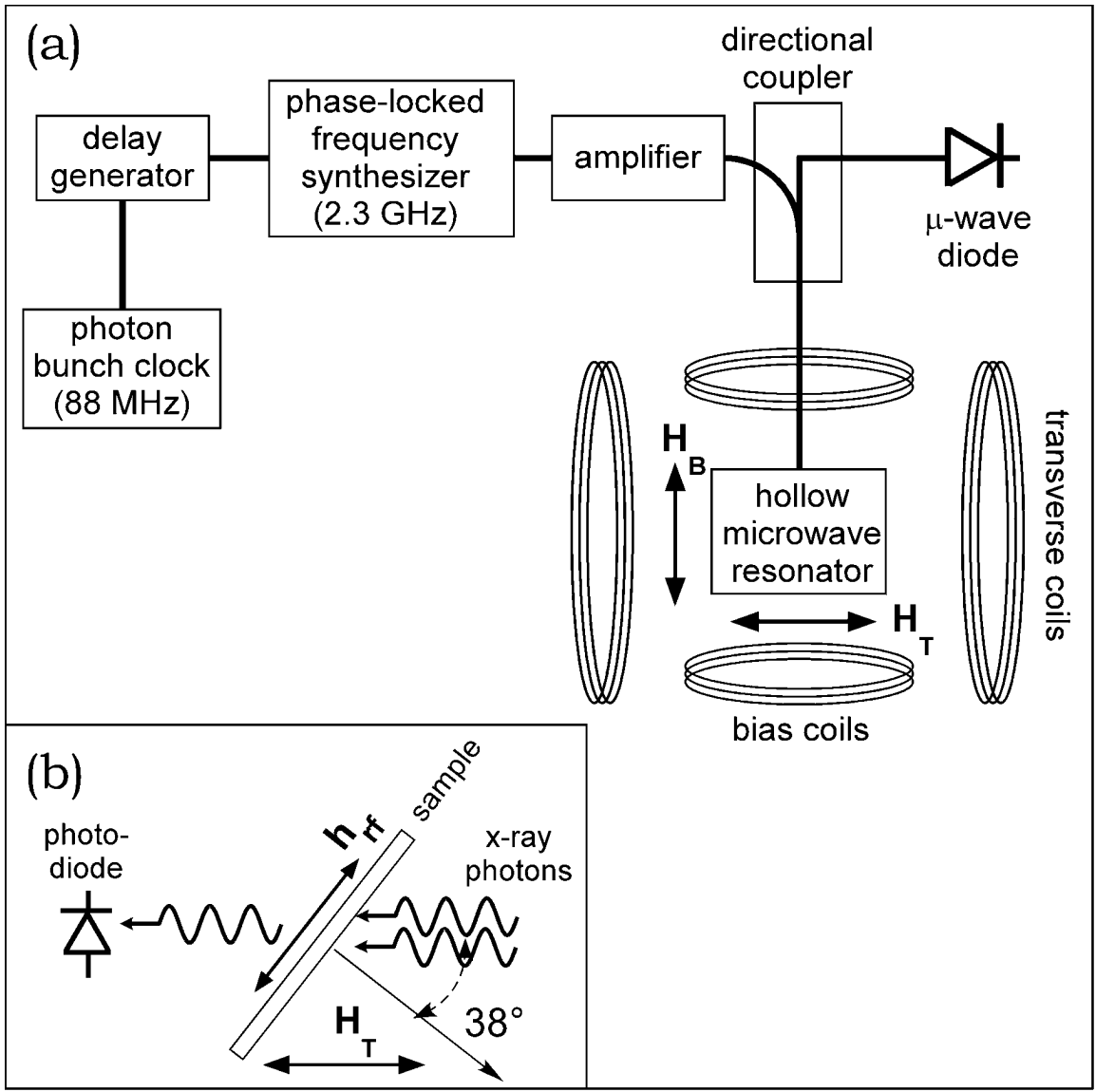

FIG. 2: Block diagram electronics used to generate phase-locked, 2.3 GHz excitation for conventional FMR and ETR-FMR measurements. (b) Top view of experimental geometry.

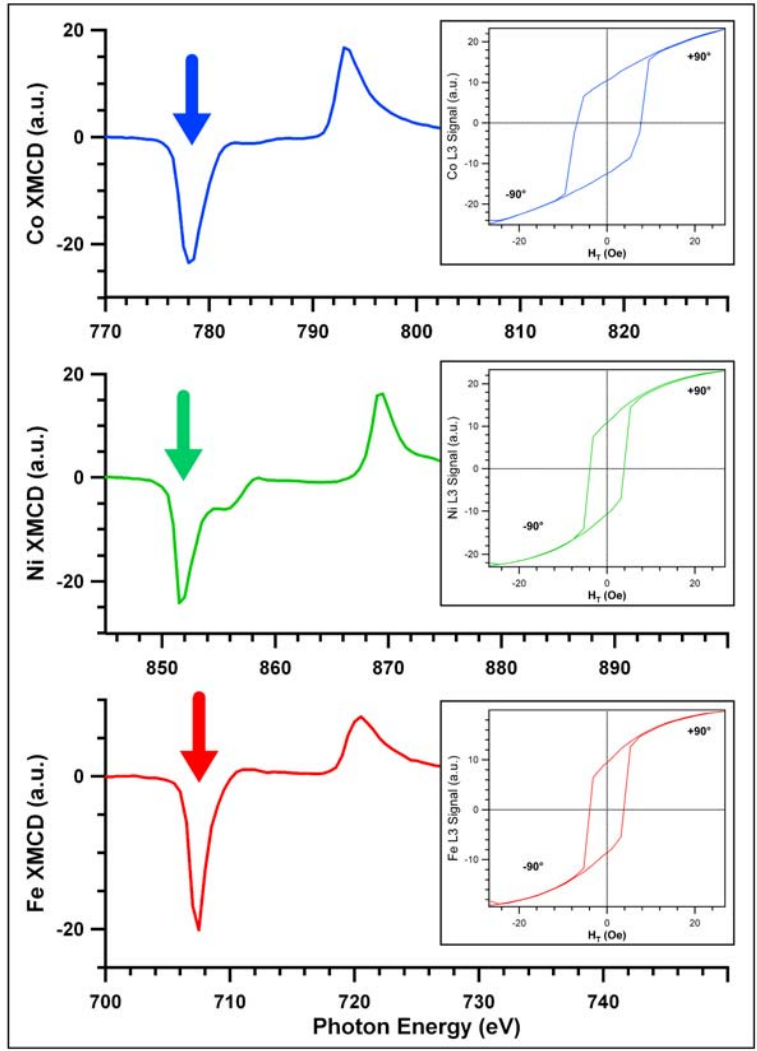

Fig. 3 (color online): XMCD spectra and element specific hysteresis measurements of the $Ni_{81}Fe_{19}$ / Cu / $Co_{93}Zr_7$ / trilayer. The hysteresis curves and ETR-FMR were measured at the photon energies indicated by the arrows (707.5 eV for Fe, 851.5 eV for Ni and 778 eV for Co). The XMCD signal levels at saturation in the hysteresis curves provide an angular calibration for ETR-FMR measurements.

-

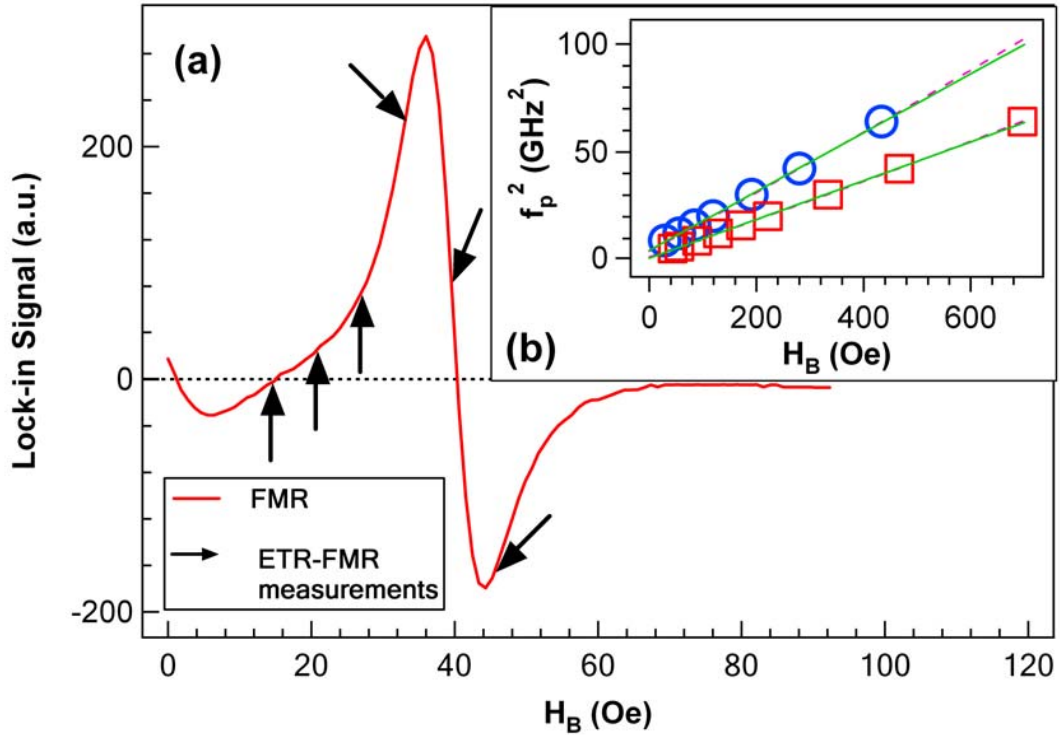

FIG. 4 (color online): (4a) *In-situ* FMR measurement at 2.3 GHz. The RF excitation used in the measurement is phase-locked with the synchrotron photon bunch clock at 88 MHz. Note the main resonance indicated by the zero crossing point at $H_B$ ~40 Oe and a weaker resonance at near zero-bias. ETR-FMR measurements were acquired at the field values indicated by the arrows. (4b) Position of resonance field $H_{B,res}$ measured with *ex-situ*, variable frequency FMR. Two resonances are observed in the trilayer, derived primarily from the $Ni_{81}Fe_{19}$ layer (red squares) and the $Co_{93}Zr_7$ layer (blue circles). See text for description of dashed / solid lines.

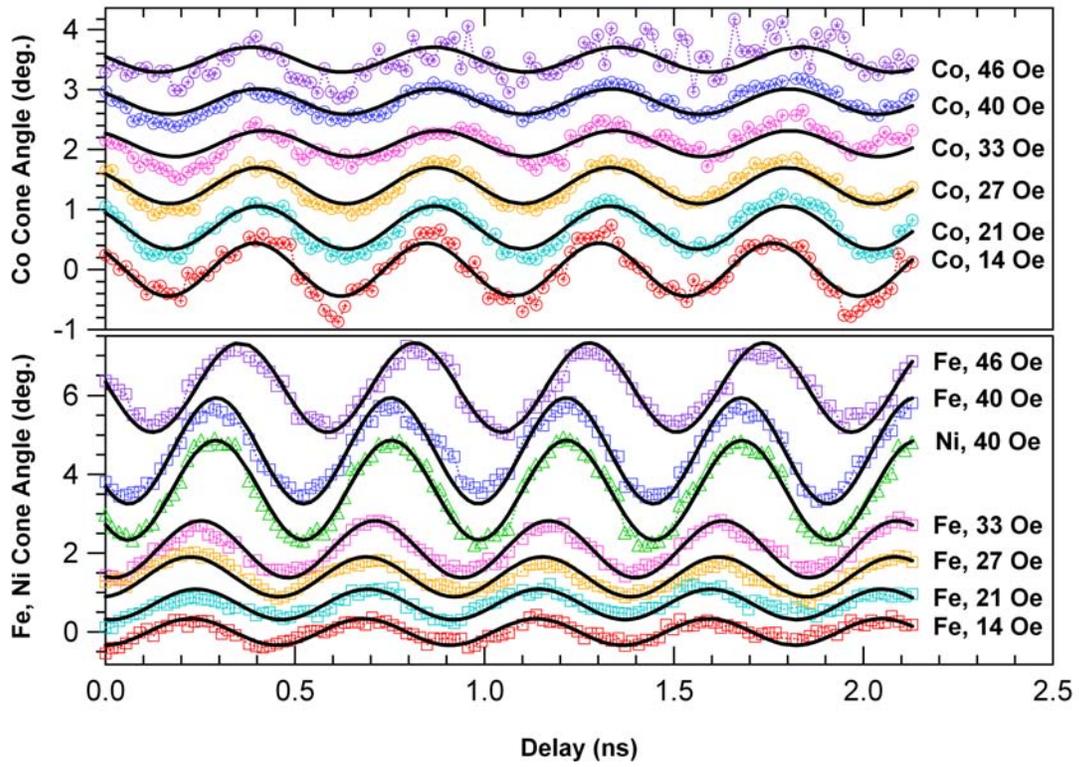

FIG. 5: (color online) ETR-FMR measurements for Fe, Ni and Co in the $Ni_{81}Fe_{19}$ / Cu / $Co_{93}Zr_7$ magnetic trilayer. The open symbols are the measured points and the solid black lines are sinusoidal fits to the data. For clarity, the data have been offset vertically.

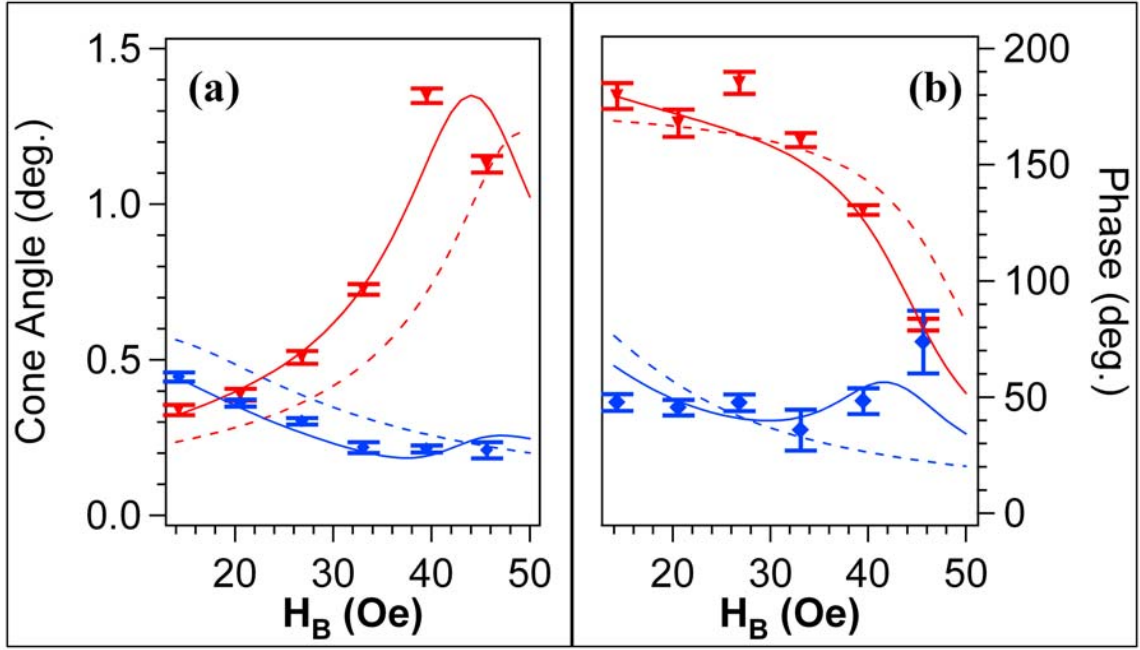

FIG. 6 (color online): Measured values for the precession cone angle (left panel) and phase of oscillation (right panel) for the Fe (red triangles) and Co (blue diamonds) moments in the $Ni_{81}Fe_{19}$ /Cu/ $Co_{93}Zr_7$ trilayer. The dashed lines are calculated values for the amplitude and phase assuming no coupling ($A_{ex} = 0$) between the layers. Solid lines are model calculations that assume weak coupling ($A_{ex} = 0.01$ ergs / cm$^2$) between the two FM layers.